\documentclass{PoS}

\usepackage{amsmath,amsfonts,amssymb,epsfig}
\usepackage{slashed}
\numberwithin{equation}{section}

\usepackage{graphicx}
\usepackage{cancel}
\usepackage{color}
\usepackage{multirow}
\usepackage{xfrac}
\usepackage{xspace}

\def\smallS{{\scriptscriptstyle S}}
\def\MS{M_\smallS}
\def\micrOMEGAs{{\tt micrOMEGAs}}

\newcommand{\scr}[1]{\ensuremath{\mathcal{#1}}}
\def\WDM{ \tilde{W} |_{\text{DM}}}
\def\BHDM{ \tilde{B}/\tilde{H} |_{\text{DM}}}
\def\HDM{ \tilde{H} |_{\text{DM}}}

\def\sp{\ \! \!}

\def\s{\sigma}
\def\d{\partial}
\def\ds{\slashed{\partial}}

\def\ks{\slashed{k}}
\def\qs{\slashed{q}}
\def\ps{\slashed{p}}

\def\g{{{\gamma}^{\sp}}}

\newcommand{\half}{  \! {\sfrac{1}{2}}}
\newcommand{\thalf}{ \! \sfrac{3}{2}}

\def\psib{\bar{\psi}}
\def\epb{\bar{\ep}}

\def\psih{{\psi_{ \sfrac{1}{2}}}}

\def\eL{\epsilon_{ \! \! \! \! \textrm{ \ \tiny LV}}}
\newcommand{\ep}{\ensuremath{\epsilon}}

\def\l{\lambda}

\def\L{\mathcal{L}}
\def\eps{\varrho}

\def\PS{\ensuremath{r}}
\def\PT{\ensuremath{t}}
\def\P{\mathcal{P}}

\newcommand{\ov}[1]{\overline{#1}}

\title{Off-trail SUSY}

\ShortTitle{Off-trail SUSY}

\author{\speaker{Karim Benakli}%
         \\
        {\em Sorbonne Universit\'es, UPMC Univ Paris 06, UMR 7589, LPTHE, F-75005, Paris, France \\}
{\em CNRS, UMR 7589, LPTHE, F-75005, Paris, France \\}
\\
        E-mail: \email{kbenakli@lpthe.jussieu.fr}}
\author{Luc Darm\' e%
         \\
       {\em Sorbonne Universit\'es, UPMC Univ Paris 06, UMR 7589, LPTHE, F-75005, Paris, France \\}
 {\em CNRS, UMR 7589, LPTHE, F-75005, Paris, France \\}
\\
        E-mail: \email{darme@lpthe.jussieu.fr}}

\abstract{We present two distinct topics: \\
I) We describe the propagation of a spin-3/2 state  in a  background which preserves invariance under space translations and rotations but not under Lorentz boost transformations. We start by building a generalisation of the Volkov-Akulov Lagrangian for a goldstino in a fluid. A super-Higgs mechanism leads to the modified Rarita-Schwinger Lagrangian describing a slow gravitino. We identify the physical propagating degrees of freedom and derive the corresponding equations of motion. This includes some new results.\\
II) Fake Split Supersymmetry Models are proposed to alleviate some of the problems of the original Split SUSY. In particular it is no more necessary to restrict to a Mini-Split scenario as higher values of the supersymmetry breaking scale (Mega-Split) are now allowed. The FSSM relies on swapping the higgsinos  for new states in identical gauge group representations but different Yukawa couplings.}

\FullConference{18th International Conference From the Planck Scale to the
Electroweak Scale \\
		 25-29 May 2015\\
		 Ioannina, Greece }

\begin{document}

\section{Introduction}
We discuss here two topics. The first is an attempt to combine Lorentz symmetry breaking  with supersymmetry in flat space-time. It is presented in section 2. The second  concerns the possibility with having the largest possible SUSY breaking scale and it is the subject of section 3. %Both subjects are departures from the usual SUSY marked tracks, hence the title. 

\section{The slow gravitino and super-Higgs mechanism in fluids}

The first part deals with study of a modified spin-3/2 Rarita-Schwinger Lagrangian~\cite{Benakli:2013ava} with two parameters, one dimension-full for the mass and the other dimensionless parameterising  the Lorentz-symmetry breaking. This Lagrangian has a rich structure and can be studied per se as done in \cite{Benakli:2014bpa}.  The gravitino exhibits an interesting feature: its longitudinal mode ($\pm 1/2$ helicities) has an effective speed of light smaller than the one appearing in the equation of motion of the transverse modes ($\pm 3/2$ helicities); thereof we call it {\it  slow gravitino}. This phenomenon is common in spin-3/2 propagation on curved backgrounds \cite{Kallosh:1999jj,Giudice:1999yt,Giudice:1999am,Schenkel:2011nv,Kahn:2015mla}.

Extensions of massive spin-3/2 Lagrangian that are free of the Velo-Zwanziger problem~\cite{Velo:1969bt} (i.e loss of causality) can be constructed through a super-Higgs mechanism. This is the case for our Lagrangian: it describes  the generation of a mass term for a goldstino appearing when SUSY is broken by a fluid background.   

Supersymmetry is broken by temperature \cite{Das:1978rx} and is not restored at any non-zero value. This is not surprising given that the bosons and fermions obey different statistics. However, the spontaneous or explicit nature of the breaking remained subject of confusion. Based on the fact that fermions have anti-periodic boundary conditions in the imaginary time, it was argued that there is no massless fermionic mode and that supersymmetry was explicitly broken \cite{Girardello:1980vv}. However, working in the real time formalism, it was found in \cite{Tesima:1982dz,Boyanovsky:1983tu} that there is a massless goldstino and supersymmetry is broken spontaneously. These results seem conflicting and have lead to a lot of debate in the literature. It was later traced back to two problems with the imaginary-time formalism: the first is that the study of linear response requires an analytical continuation to Minkowski space that is difficult to perform in the Fourier space and the second is that the imaginary-time breaks explicitly Lorentz covariance. Working in the real time formalism, it was possible to establish a Ward identity \cite{Boyanovsky:1983tu}, to show the presence of a Goldstone fermion excitation and to identify the latter in certain cases with composite bilinear boson-fermion operators~\cite{Aoyama:1984bk}. Later, the temperature goldstino has been called phonino~\cite{Lebedev:1989rz} and has been subject to further studies (see for example \cite{Leigh:1995jw,Kratzert:2003cr,Kratzert:2002gh,Kovtun:2003vj}). The only feature of the phonino that is relevant for this work is that it has a non-relativistic kinetic term, dressed by the stress-energy tensor of the fluid \cite{Hoyos:2012dh}, which involves the derivative $T^{\mu \nu} \g_\mu \d_\nu$. The phonino is expected to acquire a mass through its interaction with the  gravitino.

\subsection{Alkulov-Volkov lagrangian for the phonino}

The usual (Lorentz invariant) Alkulov-Volkov (AV) Lagrangian is given by:
\footnote{
We used the mostly plus signature $(-,+,+,+)$. The Clifford algebra is given by $\left\{ \g^a , \g^b \right\} = 2 \eta^{a b}$. Our basis for the Clifford algebra is obtained by adding to $\g^{a}$ the two matrices $\g^{ab}$ and $\g^{abc}$ defined by 
\begin{align*}
 \g^{ab} = \frac{[\g^a,\g^b]}{2} &&&
 \g^{abc} = \frac{ \lbrace \g^a,\g^{bc} \rbrace }{2} \ .
\end{align*}
We define the $\g^5$ gamma matrices with $ \g^5 = i \g^0 \g^1 \g^2 \g^3 $ and have  $
 \g^{abc} = i \epsilon^{abcd} \g^5 \g_d \ .  $} %We introduce the AV Lagrangian
\begin{equation}
\label{AVLag}
 \L ~= - \frac{1}{2} f^2 \ \det (W_\mu^{\phantom{\mu}  \nu})
\end{equation}
where $f$ is a constant parameter and
\begin{equation*}
 W_\mu^{\phantom{\mu}\nu} ~=~ \delta_\mu^{\phantom{\mu} \nu} + \frac{\bar{G}}{f}\g_{ \mu}\d^\nu (\frac{G}{f}) \ .
\end{equation*}
This Lagrangian has been chosen such that under the supersymmetry transformation of constant parameter $\ep$
\begin{align}
\label{AVtransfo}
\delta(\frac{G}{f}) ~=~ \epsilon+ \xi^\mu \d_\mu(\frac{G}{f}) &&  \textrm{ with } && \xi^\mu ~=~  \epb \g^{ \mu} \frac{G}{f} \ ,
\end{align}
we have $
 \delta( W_\mu^{\phantom{\mu} \nu}) = \d^\nu \xi_\rho W_{\mu}^{\phantom{\mu} \rho} + \xi^\rho \d_\rho W_{\mu}^{\phantom{\mu} \nu} \ ,
$ 
and 
\begin{align*}
 \delta( \L )~=~ -\frac{1}{2} f^2 \d_\mu [\xi^\mu \det (W_\rho^{\phantom{\s} \s} )] \ .
\end{align*}
The Lagrangian~(\ref{AVLag}) is therefore invariant up to total derivatives under the non-linear transformation~(\ref{AVtransfo}). Expending~\eqref{AVLag} at first order in $G^2$, we obtain 
%a constant term together with the goldstinos kinetic term 
\begin{align}
\L ~=~ -\frac{1}{2} f^2  -\frac{1}{2} \bar{G} \g^{ \mu}\d_\mu G + \scr{O} (G^4) \ .
\label{LIgolstino}
\end{align}
The main new feature of the phonino is  that instead of \eqref{LIgolstino}, the kinetic term has a non-Lorentz-invariant form $T^{\mu \nu} \g_\mu \d_\nu $. Defining $T$ or normalisation factor, the non-linear Lagrangian yielding this is given by:
%[[We need a definition for $T$ which was given in the text commented out]]
%such a kinetic term straightforwardly by making the replacement
%\begin{equation*}
% \g^\mu ~\rightarrow~ -\frac{T^{\mu \nu} \g_\nu}{T}  \ ,
%\end{equation*}
%where $T$ is a positive normalisation factor. The minus sign is necessary to ensure that with we recover the previous case when the fluid becomes a negative cosmological constant. Replacing in~\eqref{AVLag}, we obtain the following Lagrangian :
\begin{equation}
\label{PhoninoAVLag}
 \L ~=~ - \frac{1}{2} f^2 \ \det \left[ \delta_\mu^{\phantom{\mu} \nu} - \frac{\bar{G}}{f} \frac{T_{\mu \rho} \g^{ \rho}}{T}\d^\nu (\frac{G}{f}) \right] =  -\frac{1}{2} f^2  + \frac{1}{2} \bar{G} \frac{T^{\mu \nu}}{T}\g_{ \mu} \d_\nu G + \scr{O} (G^4)  \ ,
\end{equation}
with the modified non-linear supersymmetry transformations,
\begin{align}
\label{phoninoAVtransfo}
\delta(\frac{G}{f}) ~=~ \epsilon+ \xi^\mu \d_\mu(\frac{G}{f}) &&  \textrm{ with } && \xi^\mu ~=~  -\epb \frac{T^{\mu \rho} \g_{ \rho}}{T} \frac{G}{f} \ .
\end{align}
It is straightforward to  verify that~\eqref{PhoninoAVLag} transforms as a total derivative under~\eqref{phoninoAVtransfo}.

\subsection{Super-Higgs mechanism in a fluid with curved background}

\subsubsection{Perfect fluid in curved background}

To leading order in derivatives, the supergravity Lagrangian including the graviton, gravitino and goldstino fields is given by
\begin{align}
\label{LSUGRA1}
\L ~=~ \frac{1}{2} e \left[ \L_\textrm{gravitino} +  \L_\textrm{phonino} + \L_\textrm{mixing} \right] +  \L_\textrm{f}\ ,
\end{align}
where $e$ is the square root of the metric determinant and
\begin{align*}
 \L_\textrm{gravitino} & ~=~  R - \psib_\mu \g^{  \mu \nu \rho} \nabla_\nu \psi_\rho \\
\L_\textrm{phonino} & ~=~  \bar{G} \frac{T^{\mu \nu}}{T}\g_\mu \nabla_\nu G   \\
\L_\textrm{mixing} & ~=~ - \sqrt{2}\frac{T^{\mu \nu}}{\sqrt{T} M_P } \bar{G} \g_\mu \psi_\nu \ .
\end{align*}
This Lagrangian transforms as a total derivative under the linear  transformations  :
\begin{align}
\begin{cases}
 \delta e_\mu^a &=~ \frac{1}{2} M_P \bar{\ep } \g^{ a} \psi_\mu \\
 \delta \psi_\mu &=~  M_P  \nabla_\mu \ep    \\
\delta G&=~ \frac{\sqrt{T}}{\sqrt{2}} \ep \ ,
\end{cases} 
\end{align}
provided that we have 
\begin{align*}
- 2 \delta ( \L_\textrm{f}) ~=~ - M_P \epb \g_\mu \psi_\nu [\frac{1}{M_P^2} T^{\mu \nu}] \ .
\end{align*}
We show now that this is the case if  $\L_\textrm{f}$ describes  an irrotational relativistic fluid (see~\cite{Jackiw:2004nm} for a nice review). Let's denote by $\theta$ the scalar potential, $\alpha$, $\beta$ the Gauss potentials and $j^\mu$ the fluid density current. We can construct the particle number density $n$ as $$ n ~=~ \sqrt{- g_{\mu \nu} j^\mu j^\nu} \ ,$$ so that the fluid four-velocity $u^\mu$ satisfying $g_{\mu \nu } u^\mu u^\nu = -1$ is obtained from $j^\mu \equiv n \ u^\mu$.

We introduce a function $F(n)$ such that  the (perfect) fluid energy density $\eps$ and pressure $p$ are given as functions of $n$ by $\eps = F(n)$ and $p = n F'(n) - F(n)$. Then, the fluid Lagrangian can be written as:
\begin{equation}
\label{Lfluid}
 \L_\textrm{f} ~=~ - \frac{e}{M_P^2}\left[ j^\mu (\d_\mu \theta + \alpha \d_\mu \beta) + F(n)\right]
\end{equation}
in which only $n$ is a function of the metric. The Lagrangian \eqref{Lfluid} is such that combined with the Einstein term $\frac{e}{2}R$ and using the equations of motion for\footnote{A priori, we should also include the contributions from phoninos and gravitinos. However we will always suppose that their contributions to the pressure and energy density are negligible so we can use $$ \d_\mu \theta + \alpha \d_\mu \beta = j_\mu \frac{F'(n)}{n} \ . $$ We have taken the convention that the pressure is negative to match the case of the cosmological constant. For a normal fluid, one should simply takes absolute value to define $f$.} $j^\mu$ in the one for $g^{\mu \nu}$, we obtain the Einstein equations:
\begin{equation}
 R_{\mu \nu} - \frac{1}{2} g_{\mu \nu }R ~=~ \frac{T_{\mu \nu}}{M_P^2} ~\equiv~ \frac{1}{M_P^2} \left[ p   g_{\mu \nu}+ (\eps  +p)u_{\mu} u_{\nu} \right] \ ,
\end{equation}
with on the r.h.s we recognise the stress energy tensor of our perfect fluid. Notice that after integrating out $j^\mu$, we have
\begin{equation}
 \L_\textrm{f} ~=~ - \frac{e}{M_P^2} \left[j^\mu (\d_\mu \theta + \alpha \d_\mu \beta) + F(n)\right] ~=~  \frac{e p }{M_P^2}   \ .
\end{equation}
The variation w.r.t $\alpha$ and $\beta$ leads to current continuity equations describing the internal dynamics of the fluid.

The final Lagrangian~(\ref{LSUGRA1}) includes a ``source'' which will generate for instance a Friedmann-Robertson-Walker (FRW) background metric if the fluid is at rest. A similar Lagrangian has been obtained by~\cite{Kahn:2015mla} using a constrained multiplet ~\cite{Komargodski:2009rz} in a minimal effective field theory for supersymmetric inflation. In that case the fluid stress energy tensor was represented by a scalar field with $p = \frac{1}{2} \dot{\phi}^2 - V(\phi)$. In the limit where the fluid is simply a cosmological constant, we recover the case ~\cite{Bergshoeff:2015tra} that~(\ref{LSUGRA1}) can be embedded in a de Sitter supergravity action. 

\subsubsection{Super-Higgs mechanism in a fluid}

We can construct an action describing the coupling of the phonino to a gravitino in a flat space-time from~(\ref{LSUGRA1}). We start by adding to the Lagrangian above \footnote {We assume that "the fluid" does not curve space-time. An exact cancellation of the energy-momentum tensor due to a perfect fluid can be obtained by the addition of an appropriate
cosmological constant and an point-like orientifold for example. We shall not discuss these constructions.} 
\begin{align*}
-{\L}_\textrm{f} + \frac{e}{2} \left[  \frac{1}{2} \psib_\mu \g^{\mu \nu \rho} n_{\nu \l} \g^\l \psi_\rho -  \frac{T^{\mu \nu}}{2T} n_{\mu \nu } \bar{G} G\right] \ .
\end{align*}
This is   invariant under modified supergravity transformations obtained by replacing $\nabla_\mu \rightarrow \nabla_\mu - \frac{1}{2} n_{\mu \nu } \g^\nu$ under the condition that:  
\begin{align}
\label{SUSYcond}
\g^{\mu \nu \rho} ~n_{\nu \lambda} ~n_{\rho \s} \g^{ \lambda} \g^{ \s} ~=~ \displaystyle \frac{2}{ M_P^2} \tilde{T}^{\mu \nu}\g_\nu \ .
%\g^{\mu \nu \rho} [ \Omega_\nu N_\rho +  N_{\nu} \Omega_{\rho } + \del_\nu N_{\rho}  ]= \g^{\mu \nu \rho} \nabla_\nu (N_\rho) \ ,
\end{align}
The total Lagrangian is given by: 
\begin{align}
 \label{Lsugra2}
\L = \frac{1}{2} e \left[~  R  + \bar{G} \frac{T^{\mu \nu}}{T} \g_\mu \nabla_\nu G \right. & - \sqrt{2}\frac{T^{\mu \nu}}{\sqrt{T} M_P } \bar{G} \g_\mu \psi_\nu - \psib_\mu \g^{\mu \nu \rho} \nabla_\nu \psi_\rho   \\
 &\left. +\frac{1}{2} \psib_\mu \g^{\mu \nu \rho} n_{\nu \l} \g^\l \psi_\rho -  \frac{T^{\mu \nu}}{2 T} n_{\mu \nu } \bar{G} G ~\right] \ . \nonumber
\end{align}
with the supergravity transformations
\begin{align}
\label{SUGRA2}
\begin{cases}
 \delta e_\mu^a &=~  \frac{1}{2} M_P \bar{\ep } \g^a \psi_\mu \\
 \delta \psi_\mu &=~  M_P  (\nabla_\mu \ep -  \frac{1}{2} n_{\mu \nu } \g^\nu \ep ) \\
\delta G&=~ \frac{\sqrt{T}}{\sqrt{2}}\ep \ .
\end{cases} 
\end{align}
Notice that~\eqref{Lsugra2} is invariant under~\eqref{SUGRA2} only if we neglect derivatives in the fluid variables. A fully invariant Lagrangian can be nonetheless obtained even without neglecting them~\cite{Benakli:2013ava}. In the unitary gauge obtained by
\begin{align*}
\begin{cases}
 \delta e_\mu^a &=~ - \frac{1}{\sqrt{2 T}} M_P \bar{G} \g^a \psi_\mu \\
 \delta \psi_\mu &=~  - M_P  (\nabla_\mu  -  \frac{1}{2} n_{\mu \nu } \g^\nu  ) \frac{\sqrt{2} G}{\sqrt{T}} \\
\delta G &=~ -G \ ,
\end{cases} 
\end{align*}
the phonino is removed from the Lagrangian that becomes
\begin{align}
\label{lagrangiangeneral}
 \L = \frac{1}{2} e  & \left[~  R - \psib_\mu \g^{\mu \nu \rho}  (\nabla_\nu - \frac{n_{\nu \l} \g^{ \l}}{2})  \psi_\rho ~\right]
\end{align}
where $ n_{\nu \l}$ satisfies~(\ref{SUSYcond}). The gravitino acquires a mass similarly to the usual super-Higgs mechanism. A distinct feature, however, is that the mass terms is now space-time dependent and violates  Lorentz invariance when $n_{\mu \l}$ is not proportional to $g_{\mu \nu}$.

%With the effect of the fluid integrated in the graviton components, we can now treat this Lagrangian ``per se'' in flat space-time.

\subsection{Slow gravitino}

The Lagrangian~\eqref{lagrangiangeneral} describes a ``slow gravitino''.  We present here some of its properties.  More details can be found in~\cite{Benakli:2014bpa}.

We will consider a perfect fluid with four-velocity $u^\mu$ and an equation of state 
\begin{equation}
w~=~ \frac{p}{\eps} \ . \label{state2}
\end{equation}
For $w\neq-1$ both supersymmetry and invariance under Lorentz boosts are spontaneously broken while $w = -1$ corresponds to a cosmological constant. It can be shown~\cite{Benakli:2013ava} that supersymmetry requires the gravitino mass to be given by:
\begin{align}
\label{massgr}
m  &~=~  \frac{ \sqrt{ 3\eps }}{ 4 M_P} ~ |\frac{1}{3} - w| \ .
\end{align}
From now on, we will neglect all derivatives of the fluid variables compared to the momentum or the mass of the gravitino and, when convenient, will trade the fluid variables $\eps$ and $p$ for
\begin{align}
m, \qquad \eL ~\equiv~ 1+w , 
\end{align}
where $m$ is the mass (\ref{massgr}), and  $\eL$ measures the size of violation of Lorentz boost invariance. We also define at every point in space-time two projectors $\PS$ and $\PT$ by
\begin{align}
\begin{split}
\label{PmunuDefFirst}
& \PS^{\mu \nu} ~\equiv~ \eta^{\mu \nu} + u^\mu u^\nu \\
 & \PT^{\mu \nu} ~\equiv~ (\mathbf{1} -  \PS)^{\mu \nu} = - u^{\mu} u^{\nu} \ .
\end{split}  
\end{align}
$\PT$ projects along $u^\mu$, i.e. in the time-like direction defined by the fluid, while $\PS$ projects on the vector space orthogonal to
$u^\mu$, i.e. on the spatial vector space defined by the fluid. Since our fluid is decribed by the Lagrangian~(\ref{Lfluid}), it is irrotational and its velocity does  define a foliation of space-time that we can use to defined plane waves of the form $\psi^\mu \propto e^{i p^\mu x_\mu}$ with $p^{\mu}$ being
functions of the space-time coordinates whose derivatives are neglected.
%s, but their derivatives are neglected since they depend on fluid variables. 
This will allow us to define helicitiy eigenstates and construct the corresponding propagator.

Using the fluid foliation, we define the ``spatial'' and ``temporal'' components of the gamma matrices $\g^\mu$ and of the momentum $p^\mu$, defined via using the projectors $\PS$ and $\PT$. They
are constructed  as
\begin{align*}
r^\mu &= \PS^{\mu \nu} \g_\nu \ & k^\mu &= \PS^{\mu \nu} p_\nu \nonumber \\
 t^\mu &= \PT^{\mu \nu} \g_\nu \ & q^\mu &= \PT^{\mu \nu} p_\nu  \ .
\end{align*}
$r^\mu$ and $t^\mu$ behave as $\g^{ i}$ and $\g^{ 0}$.
They satisfy the relations $r^\mu r_\mu = 3$, $t^\mu t_\mu = 1$ and $t^\mu
r^\nu = -r^\nu 
t^\mu$. Using these objects, our Lagrangian~\eqref{lagrangiangeneral} takes the form:
\begin{align}
\label{lagrangianMaster}
 \L ~=~  \frac{1}{2}    \psib_\mu  \!   \left[~ (\g^{\mu \nu})(-\ds -  \! m )+ \!  \g^\mu \d^\nu  \! 
 - \g^\nu \d^\mu  \!  - \frac{3\eL m}{4-3\eL}  (r^\mu t^\nu + t^\mu r^\nu)
~\right]  \! \psi_\nu \ .
\end{align}
In (\ref{lagrangianMaster}) one identifies the first term with the usual Rarita-Schwinger Lagrangian and the term proportional to $\eL$ as the correction due to violation of Lorentz invariance. 

We construct a spin-$3/2$ field from the product of spin-1/2 and spin-1 states (a spinor-vector) denoted as $\psi^\mu$. This is a reducible representation of the rotation group that can be decomposed into spin representation as
\begin{align*}
\label{decompose}
(\frac{1}{2},\frac{1}{2}) \otimes (\frac{1}{2},0) =  \frac{1}{2} \oplus (1 \otimes \frac{1}{2}) = \frac{1}{2} \oplus \frac{1}{2} \oplus \frac{3}{2} \ .
\end{align*}
In general, we will therefore decompose $\psi_\mu $ into four spinors  corresponding to the helicity-$\frac{3}{2}$ states $\psi^\mu_{\thalf}$, the helicity-$\frac{1}{2}$  states $\psih$, and two remaining un-physiscal spinors that are projected out by two constraints. They are obtained from the equations of motions~\cite{Benakli:2014bpa} by contracting with either $u_\mu$ to extract the temporal part, or by the derivative operator $\nabla_\mu - N_\mu/2$ and read
\begin{equation}
\left[ r^\mu r^\nu - \PS^{\mu \nu}\right] \d_\mu \psi_\nu ~=~  - \frac{m}{1-\frac{3}{4} \eL}  r^\rho \psi_\rho \ , \label{C1}
\end{equation}
and 
\begin{equation}
(w r^\nu - t^\nu ) \psi_\nu ~=~ 0   \\    \\ \label{C2} \ ,
\end{equation}
One can use these constraints to obtain the physical degrees of freedom $\psi^\mu_{\thalf}$ and $\psih$. We can obtain them directly from $\psi_\mu$ by
\begin{align}
\psih & ~=~ \sqrt{\frac{3}{2}} \frac{m}{k(1-3\eL/4)} \slashed{u} ~ r^\rho \psi_\rho \ \\ \nonumber
\psi^\mu_{\sp \thalf} & ~=~ \P_{\thalf}^{\mu \nu} \psi_\nu ~\equiv~ \left[ \eta^{\mu \nu} - \frac{1}{3} r^\mu r^\nu - t^\mu t^\nu 
- \frac{1}{6} (r^\mu - 3 \frac{\ks k^\mu}{k^2})(r^\nu - 3\frac{\ks k^\nu}{k^2}) \right]
\psi_\nu \ .
\end{align}
Note that the spin-1/2 degrees of freedom are proportional to $r^\mu \psi_\mu$ up to a normalisation factor and a gamma matrix. These factors are crucial in obtaining a correctly normalised fermionic kinetic term for $\psih$.
 The equations of motion for these fields then take the form
\begin{eqnarray}
\label{eoms}
 (  t^\rho \d_\rho   + r^\rho\d_\rho +   m) \psi^\mu_{\sp \thalf}   &=&~ 0  \ , \nonumber\\ 
( t^\rho \d_\rho - wr^\rho \d_\rho + m ) \psih &=&~  0 \ .
  \end{eqnarray}

Another interesting object that can be derived from the Lagrangian~\eqref{lagrangianMaster} is the propagator for the gravitino. We find that it can be written as
\begin{align}
 G^{\mu \nu} &= \frac{\Pi_{~\thalf}^{\mu \nu}}{p^2 + m^2} + \frac{\Pi_{~\half}^{\mu
\nu}}{w^2 k^2 + q^2 + m^2  }  - \frac{3}{4} \eL \frac{\ks}{m k^2}(t^\mu k^\nu - 
k^\mu t^\nu) \ .
\end{align}
It has two parts corresponding to the helicity-3/2 and helicity-1/2 components of the spinor-vector. Both have different poles, thus different dispersion relations. The quantities $\Pi_{~\thalf}^{\mu \nu}$ and 
$\Pi_{~\half}^{\mu \nu}$ are the corresponding polarisations and take the form 
\begin{align*}
\Pi_{~\thalf}^{\mu \nu} = & (m - i \ps) \P_{\thalf}^{\mu \nu} \ ,
\end{align*}
and
\begin{align*}
\begin{split}
\Pi_{~\half}^{\mu \nu} =&   \frac{2}{3}\,  \,  \Lambda^\mu \,  \,  (i \ps  - \eL i \ks + m)\,  \,  \Lambda^\nu  \ , 
\end{split} 
\end{align*}
where
\begin{equation*}
\Lambda^\mu = \g^\mu -i \frac{p^\mu}{n} - \frac{3}{2} (r^\mu - \frac{\slashed k k^\mu}{k^2}) - \frac{3}{4} \eL t^\mu  \ ,
\end{equation*}
In the limit of gravitino high momentum where we have the hierarchy
\begin{align*}
m ~ \ll ~ |p| ~\ll ~ f \ ,
\end{align*}
the propagator simplifies to
\begin{align}
 G^{\mu \nu} &\rightarrow  - \P_{\thalf}^{\mu \nu} \frac{ i \ps}{p^2}   - \frac{2}{3}\frac{p^\mu p^
\nu}{n^2} \frac{i \qs -  i w\ks}{q^2 + w^2 k^2 } \ .
\end{align}

\section{ The Fake Split Supersymmetry Model}

\subsection{Constraints from the Higgs mass}

The ATLAS and CMS collaborations have discovered a Higgs particle candidate that is very much SM-like. At the lowest (renormalisable) order, the SM Higgs scalar potential is governed by two parameters. These can either be taken to be the coefficients of the quadratic and quartic terms or the Higgs vacuum expectation values and masses. We shall be using both sets here depending on which one appears to be more convenient.

At tree level,  the Higgs quartic coupling $\lambda$  is given in the MSSM by the $SU(2)\times U(1)$  $D$-term. It is corrected by radiative corrections which are known at leading orders.  For a generic supersymmetric model, there might be additional contributions that arise from the superpotential.  We are interested in finding the set of  parameters of the model such that the computed $\lambda$ is compatible with the measured Higgs mass? In particular, we would like to determine the allowed range for  $\MS$, the scale of supersymmetry breaking.

At $M_S$, we have to match an effective low energy theory for the light degrees of freedom with a supersymmetric theory that takes into account all supersymmetric partners. This matching has in particular to be done for the Higgs quartic coupling, leading to a boundary condition for the corresponding RGEs. Solving the RGEs leads to a prediction for the Higgs mass and allows to find the set of viable parameters of the theory. Such an anlysis was conducted for the case of large $M_S$  for three scenarios:
\begin{enumerate}
\item Split SUSY: Above $M_S$ we have the MSSM while below $M_S$ the spectrum is that of the Standard Model supplemented by the whole supersymmetric fermionic partners  (gauginos and higgsinos) with TeV scale masses \cite{ArkaniHamed:2004fb,GiudiceRomanino}.

\item High scale SUSY: Above $M_S$ we have the MSSM while below $M_S$ we have only the SM particles.

\item The Fake Split SUSY Models (FSSM): Below $M_S$ a particle content with quantum numbers (SM gauge group representations) similar to those of Split SUSY. But the higgsinos and (depending of the model) the gauginos are swapped with states with different Yukawa couplings as we shall discuss below. 

\end{enumerate}

It was then found in Split SUSY that a $125$ GeV Higgs can be obtained only for $\MS$ between about $10^4$-$10^8$ GeV region~\cite{ArkaniHamed:2004fb,GiudiceRomanino,ArkaniHamed:2004yi,bernal_mssm_2007,
Giudice:2011cg}. While, for High Scale SUSY the maximal $\MS$ is of order  $10^{11}$-$10^{12}$ GeV. Things get even worse when one requires unified soft masses for the Higgs fields, as this takes Split SUSY to a mini-corner of parameters: the mini-Split case with $\MS$ in the range $10^4$-$10^6$ GeV.

We would like to find  a way to rescue the original idea of Split SUSY: allowing an arbitrary value of $\MS$ and an arbitrarily large splitting in the sparticles masses. This is one aim of FSSM  ~\cite{FSSM1} and \cite{FSSM2}. There are different realisations of the FSSM scenario (see also~\cite{dudas_flavour_2013} for yet another different realisation and for a motivation of fake gluinos):
\begin{itemize}
\item  FSSM-I: both the higgsinos and gauginos are swapped for fake gauginos (henceforth F-gauginos) and fake higgsinos (henceforth F-higgsinos). 
\begin{enumerate} 
\item The fermions remain light because of a $U(1)$ flavour symmetry 
\item The F-gauginos are Dirac partners of the gauginos  
\item Higgs-F-higgsino-F-gaugino Yukawa couplings ${\tilde g}_{u,d}, {\tilde g'}_{u,d}$ are suppressed by $({\rm TeV}/M_S)^2$
\item Two pairs of vector-like electron superfields  need to be added at $M_S$ to insure unification.
\end{enumerate} 

\item FSSM-II: Only the higgsinos are swapped for fake higgsinos (henceforth F-higgsinos). 
\begin{enumerate} 
\item The fermions remain light because of R-symmetry charges
\item Higgs-F-higgsino-gaugino Yukawa couplings ${\tilde g}_{u,d}, {\tilde g'}_{u,d}$ are suppressed by  $({\rm TeV}/M_S)$
\item Two pairs of Higgs-like doublets needed for the F-higgsinos
\item Two pairs of $(\mathbf{3}, \mathbf{1})_{1/3} \oplus  (\ov{\mathbf{3}}, \mathbf{1})_{-1/3}$. In total we have added a vector-like pair of $\mathbf{5} + \ov{\mathbf{5}} $ of $SU(5)$ to insure unification.
\end{enumerate} 
\end{itemize}

The success of FSSM is illustrated in Figure~1. Technical details of the computations can be found in~\cite{FSSM1} and \cite{FSSM2}. Here, we would like to pinpoint the origin of the  differences in the predictions for the Higgs mass. While in Split SUSY, the Higgs mass regularly increases with $\MS$, the increase is way flatter in High Scale SUSY and the curve even exhibits a plateau for the FSSM.

\begin{figure}[!htbp]
\begin{center}
\includegraphics[width=0.8\textwidth]{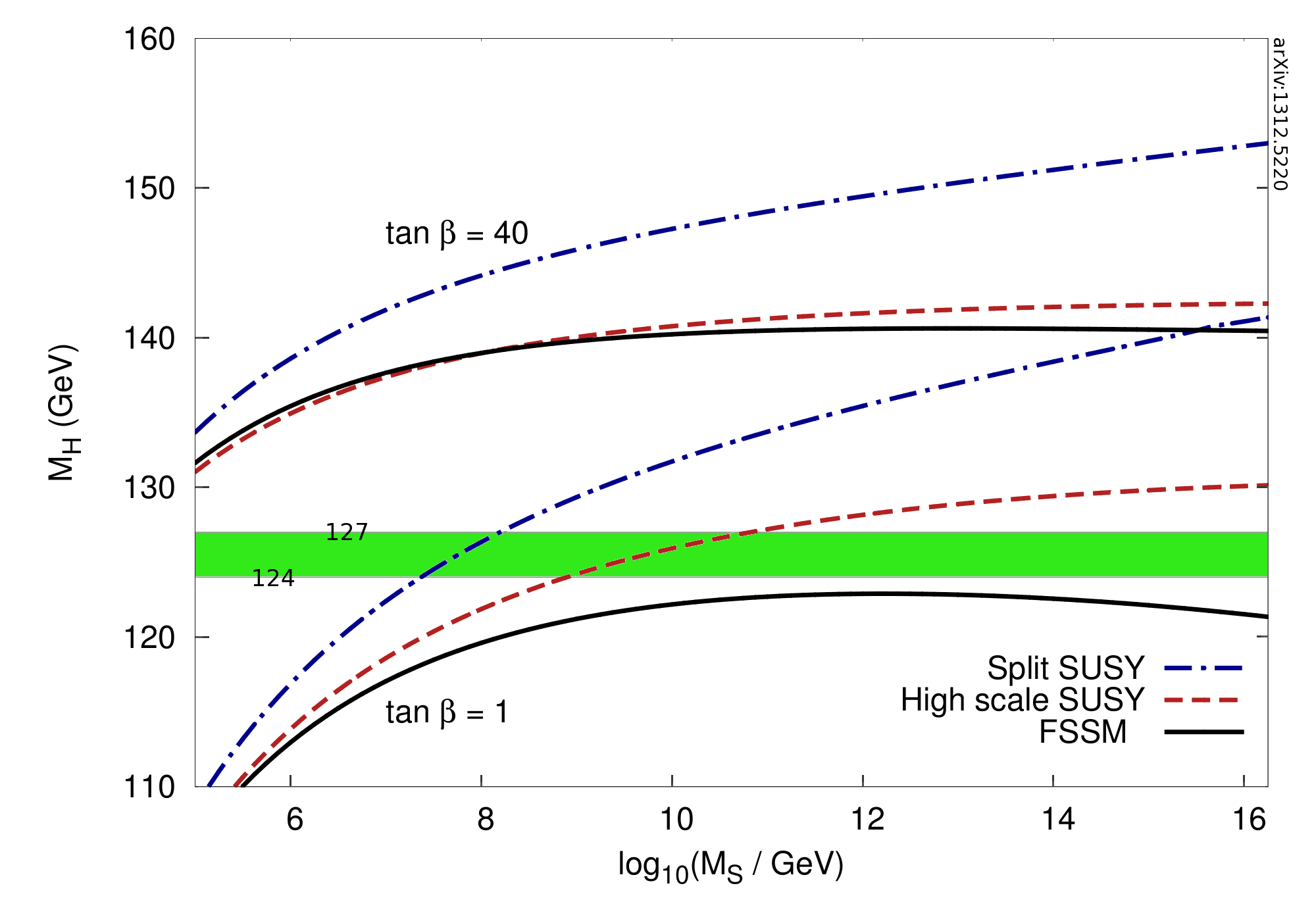}
\caption{ \footnotesize Higgs-mass predictions as a function of the
  SUSY scale $\MS$ for FSSM-I, High-Scale SUSY and Split SUSY. The green-shaded
  region indicates a Higgs mass in the range $[124,127]$ GeV. }
\vspace*{-2mm}
\end{center}
\label{fig:mhiggs_multi}
\end{figure}

Let us first describe how the Higgs mass is obtained in these models. The precise computations implies solving the RGEs through an iterative numerical algorithm which successively applies the boundary conditions at $M_S$ and at the electroweak scale. 
Schematically :
\begin{itemize}
\item We start by the determination of the values of gauge and Yukawa couplings at
$\MS$ by evolving them from the experimental values at the
electroweak scale to the SUSY scale.

\item We determine the numerical value of $\lambda (\MS)$ at the SUSY scale from
its dependance on the gauge and Yukawa couplings. In the models under
consideration $\lambda ~=~
\frac14\left(g^2+g^{\prime\,2}\right)\,\cos^22\beta + \Delta \lambda$, where $\Delta \lambda$ includes the threshold effects (suppressed in FSSM as discussed in~\cite{FSSM1})

\item We run backwards in energies, use again the experimental values at the electroweak, and iterative the previous steps until convergence.

\end{itemize}
In order to understand the hurdles met when trying to reproduce the
correct Higgs mass with an arbitrary high $\MS$, let us look at the way
$\lambda$ evolves. The various contributions to $\beta_\lambda$ at one-loop can be roughly classified as\footnote{Studying $\beta_\lambda$ at one-loop is enough to understand the two main mechanisms discriminating the three cases in Figure~1.}:
\begin{align}
\label{eq:lam}
 \beta_\lambda= \frac{1}{16 \pi^2}  & \left[  \underbrace{ 12 \lambda^2  +  \lambda( 12 y_t^2 + (\cdots \tilde{g}^2 \cdots) - (\cdots g^2 \cdots) )}_{\equiv ~\beta_{\text{quartic}}}  + \underbrace{(\cdots g^4 \cdots)}_{\equiv ~\beta_{g}}   \underbrace{- (\cdots \tilde{g}^4 \cdots)}_{\equiv ~\beta_{\tilde{g}}} \underbrace{-12 y_t^4}_{\equiv ~\beta_{t} } \right] \ ,
\end{align}

where $(\cdots g^n \cdots)$ and $(\cdots \tilde{g}^n \cdots)$ contains contributions from contains gauge couplings and Higgs-higgsino-gaugino Yukawa couplings, respectively. Fixing $\lambda$ at $\MS$ and evolving it down to the electroweak scale, positive contributions tend to bring
$\lambda$ towards lower values while negative contributions increases the Higgs mass. Two different effects explain the discrepancies between Split SUSY, High-Scale SUSY and FSSM:
\begin{enumerate}
 \item Compared to Split SUSY, we see that both FSSM and High scale SUSY have vanishing $(\cdots \tilde{g}^n \cdots)$ terms. This decreases $\beta_\lambda (\MS)$ in the Higgs scale SUSY and FSSM case.
 \item High scale SUSY has smaller gauge couplings than FSSM and Split SUSY:  these have extra fermions below $\MS$ which contribute in RGEs to push the couplings towards higher values. 
 \end{enumerate}

These corrections are  enhanced by a ``domino'' effect. At one-loop the top Yukawa coupling $y_t$ beta function $\beta_{t}$ has a positive contribution from $(\cdots \tilde{g}^n \cdots)$ terms and a  negative one from $g_3$. Since $y_t$ is fixed at the electroweak scale, a smaller $\beta_{y_t}$ means a smaller $y_t$ at $\MS$. The split between different contributions to the $\beta$-functions is presented in Figure~2.

\begin{figure}[!htbp]
\begin{center}
\includegraphics[width=0.495\textwidth]{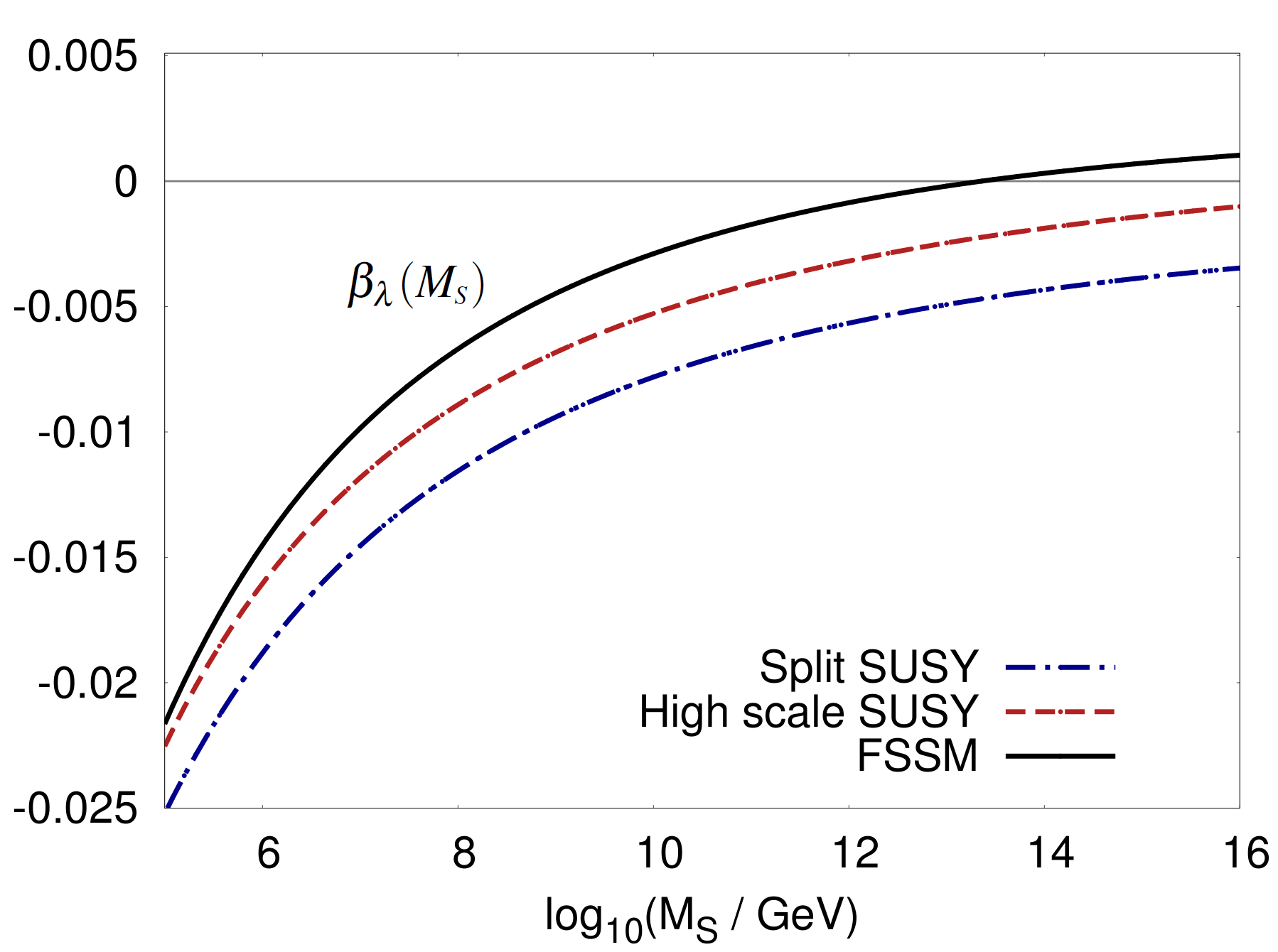}
\includegraphics[width=0.495\textwidth]{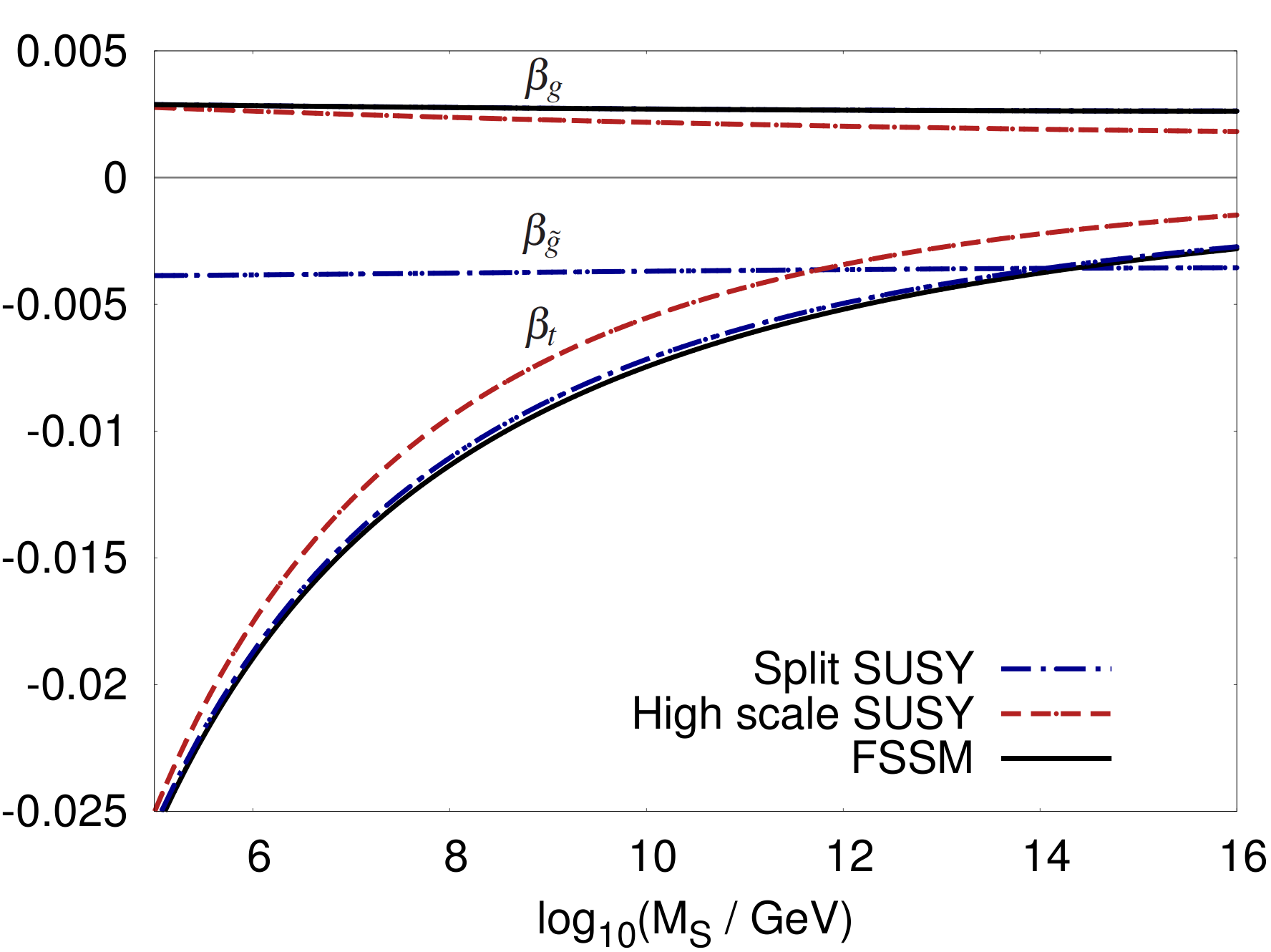}
\caption{ \footnotesize \textbf{Left plot :} $\beta_\lambda $ at $\MS$ for $\tan \beta = 1$ in the case of Split SUSY, High scale SUSY and FSSM-I as a function of $\MS$. \textbf{Right plot :} Decomposition of $\beta_\lambda $ at $\MS$ into its various components $\beta_{t}$, $\beta_{g}$ (superposed for FSSM and Split SUSY) and $\beta_{\tilde{g}}$ as a function of $\MS$.}
\vspace*{-2mm}
\end{center}
\label{fig:beta}
\end{figure}

To summarise, the success of FSSM rests on simultaneously switching off the higgsinos Yukawa couplings while conserving stronger gauge couplings than High scale SUSY thanks to the presence of extra states at the TeV-scale. Similar arguments also explain why the Extended Split SUSY scenario presented in the Appendix of~\cite{FSSM2} predicts lower Higgs mass than the three scenarios described here. It keeps the suppression of $(\cdots \tilde{g}^n \cdots)$ terms while it enhances the gauge couplings.

\subsection{Mega-Split vs Mini-Split}

Another issue \cite{Arvanitaki:2012ps} in Split SUSY is that $M_S$ is further constrained to lower values when considering unified soft masses fro the Higgs field and tuning the Higgs mass at its measured value. This comes from the fact that RGEs for the Higgs sector soft masses does not preserve their  quasi-degeneracy when running down from the unification scale. The longer the running above $M_S$, the higher the predicted  $\tan \beta$, which leads to a higher Higgs mass at tree level. Hence it is natural to have larger values of  $\tan \beta$ for small values of $M_S$, while for larger $M_S$ we can have  $\tan \beta \sim 1$. A large value for $\tan \beta$  is not compatible with a 125 GeV Higgs mass with a large $M_S$ in the case of Split SUSY models~\cite{bernal_mssm_2007,Giudice:2011cg} thereof the "Mini-Split". 

It is easy to see that low value of $\tan \beta$ are natural in FSSM, while this is not the case the Split SUSY models. At the SUSY scale, $\tan \beta$ is defined fixed by the requirement that one Higgs has a mass at the electroweak scale. In Split SUSY this fixes 
\begin{align*}
 \tan \beta ~=~ \sqrt{\frac{m_{H_d}^2 + |\mu|^2}{m_{H_u}^2 + |\mu|^2}} ~\simeq~ \sqrt{\frac{m_{H_d}^2 }{m_{H_u}^2} }
\end{align*}
since higgsinos are light. Renormalisation group evolution of $m_{H_d}^2$ and $m_{H_u}^2$, particularly due to the large top Yukawa coupling, then  generates  a $ \tan \beta \gg \scr{O } (1) $. On the contrary, in FSSM, we have
\begin{align*}
 \tan \beta ~=~ \sqrt{\frac{m_{H_d}^2 + |\mu_d|^2}{m_{H_u}^2 + |\mu_u|^2}} \ ,
\end{align*}
with $\mu_d = \mu_u$ in the FSSM-I and $\mu_u ,\mu_d \simeq \MS$ since they are unrelated to the low energy spectrum. Note that the longer the running above $M_S$, the higher the predicted  $\tan \beta$, which in turn raises the Higgs mass at tree level. Hence for small values of $M_S$ it is natural to have larger values of  $\tan \beta$, and for larger $M_S$ we expect  $\tan \beta \sim 1$.

In Figure~3 where we have plotted  $\tan \beta $ in the FSSM-I as a function of unified SUSY-breaking scalar mass $m_0$ and the A-term at the GUT scale $A_0$. We see that in most of the parameter space  $\tan \beta $ is  between $1$ and $1.4$. The increase in the right part of the plot show that for a larger value of $A_0$, $m_{H_u}^2 + \mu_0^2$ can run close to zero. In principle, by varying $m_0$ and $A_0$ in the FSSM-I we can find values of  $\tan \beta  > 2$,  potentially allowing values of the SUSY scale lower than $10^9$ GeV without requiring a breaking of the universality of the soft masses at the GUT scale.

\begin{figure}[!htbp]
\begin{center}

\includegraphics[height=0.38\textheight]{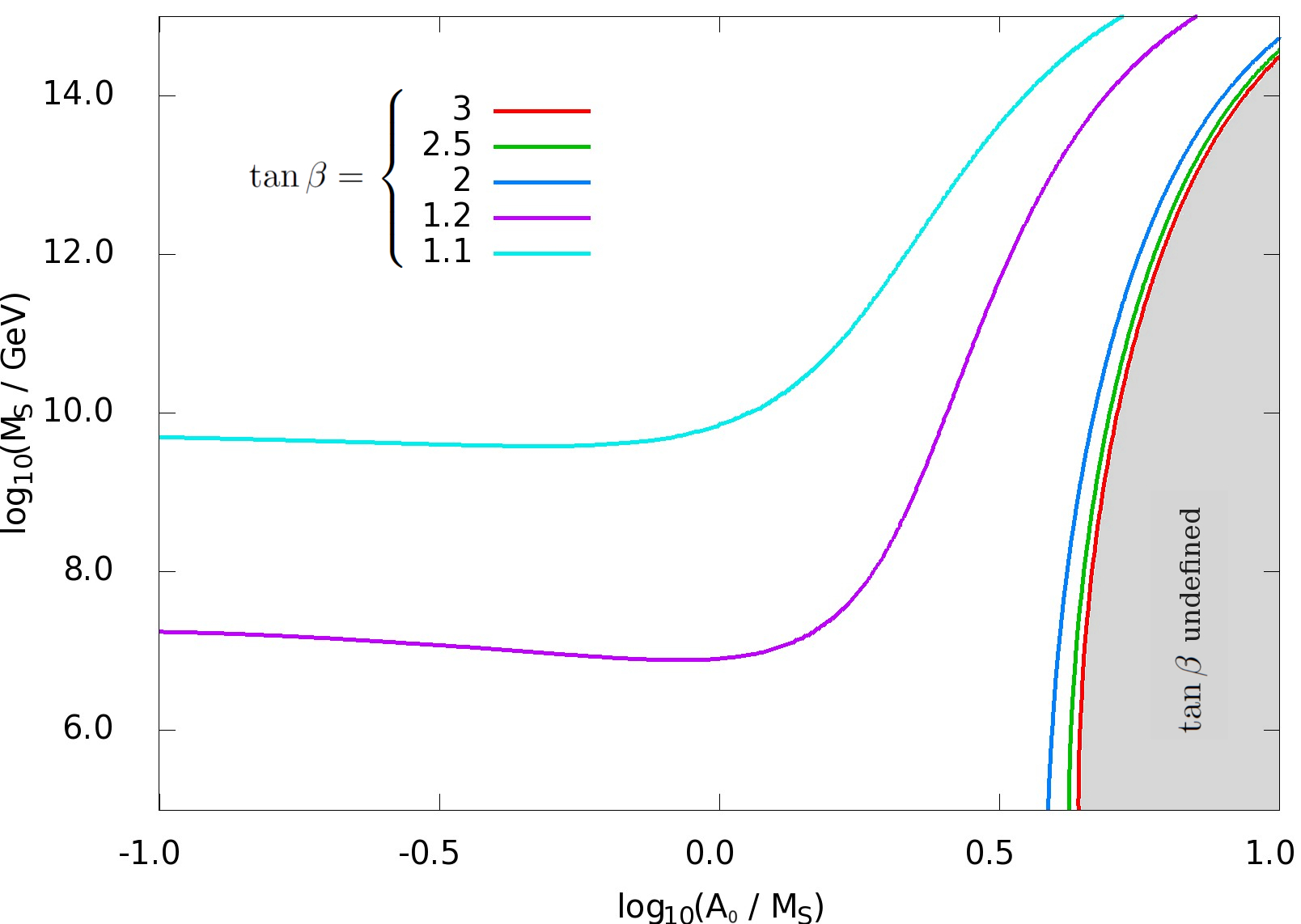}
\caption{  Contours of the value of $\tan \beta  = \sqrt{\frac{m_{H_d}^2 + |\mu_d|^2}{m_{H_u}^2 + |\mu_u|^2}}  $ in the FSSM-I varying the SUSY scale $\MS$ and the trilinear mass at $\MS$, $A_0$. }
\label{fig:tanb_at_m0}
\end{center}
\end{figure}

\subsection{Constraints from cosmology}

It is a standard result that gluinos are long-lived in SUSY models with heavy $\MS$. Since F-gluinos decay must proceed via mixing with the usual gluinos, their decay rates are even more suppressed. In~\cite{FSSM1,FSSM2} we show that $ \tau_{\tilde g'}$, the F-gluino life-time is given by
\begin{align}
 \tau_{\tilde g'}  ~\sim~&
  4 \ \text{sec}\times\left(\frac{M_S}{10^7\text{GeV}}\right)^6
\times\left(\frac{1     \ \text{TeV}}{m_{fg}}\right)^7~ ,
\end{align} 
in the FSSM-I, where $m_{fg}$ is the F-gauginos mass scale. While in the FSSM-II, since the gauginos are not fake, this enhancement does not occur and one is left instead  with the Split SUSY gluino life-time
\begin{align}
 \tau_{\tilde g'}  ~\sim~&
  4 \ \text{sec}\times\left(\frac{M_S}{10^9\text{GeV}}\right)^4
\times\left(\frac{1     \ \text{TeV}}{M_{1/2}}\right)^5~ ,
\end{align} 
where $M_{1/2}$ is the gauginos mass scale.

If one sticks with a standard cosmology the (F)-gluino lifetime is severely constrained, $\tau_{\tilde g} < 100\text{s}$~\cite{arvanitaki_limits_2005}.  Big-Bang Nucleosynthesis (BBN), the CMB spectrum and the gamma-ray background ruled out relic (F)-gluinos with lifetime between $10^2$ s until $10^{17}$ s. When the (F)-gluino is stable at the scale of the age of the universe, heavy-isotope searches also rule out such relic (F)-gluinos. This translates into limiting the SUSY scale to be below $5 \cdot 10^8$ GeV for the FSSM-I and $5 \cdot 10^{10}$ GeV for the FSSM-II. As it was underlined in~\cite{FSSM1}, these constraints depends on wether or not one considers a ``standard'' cosmology. A late time reheating occuring before BBN could for instance dilute gluino relic. In such case, heavy-isotope searches are so stringent that they still constraint $\tau_{\tilde g} \lesssim  10^{16} \text{s}$ but one can avoid constraints from the CMB spectrum and the gamma-ray background, allowing therefore SUSY scales up to $10^{10}$ GeV for the FSSM-I and $ 10^{14}$ GeV for the FSSM-II.

It is however easy to find viable dark matter candidates in FSSM. We have distinguished three scenarios : 
\begin{itemize}
 \item Scenario $\HDM$: F-higgsino LSP.
 \item Scenario $\WDM$: (F-)Wino LSP.
 \item Scenario $\BHDM$: a mixed F-Bino/F-higgsino LSP, with a small splitting.
\end{itemize}
The relevant constraints are summarised in Table~\ref{table:constraints}.
\begin{table}[h]
\centering
\begin{footnotesize}
\begin{tabular}{|l||p{3.5cm}|l|l|}
  \hline
 \textbf{DM type} & \textbf{Inelastic scattering} & \textbf{Relic density} & \textbf{Gluino lifetime} \\[0.5em]  \hline  
\rule{0pt}{3ex}  \textbf{$\mathbf{\WDM}$}    & None &  $m_{\tilde{W}} \subset [2390,2450]$ GeV  & \multirow{3}{3.7cm}{For multi-TeV  gluinos  \\ $ \begin{cases}
          M_S \lesssim 5 \cdot 10^{8} \text{GeV} \\  \text{(for FSSM-I)} \\ \\
	  M_S \lesssim  2 \cdot 10^{10} \text{GeV} \\  \text{(for FSSM-II)}
         \end{cases} $ } \\[0.8em] \cline{1-3}
\rule{0pt}{3ex}   \textbf{$\mathbf{\BHDM}$} & $\mu_\text{pole}  \lesssim 900$ GeV& $m_{\tilde{B}} \simeq \mu_\text{pole}  - (900 - \mu_{\rm pole})/x_f$&  \\[0.3em] \cline{1-3} 
\rule{0pt}{3ex}  \textbf{$\mathbf{\HDM}$} &  $ \begin{cases}
          M_S \lesssim 5 \cdot 10^{6} \text{GeV} \\  \text{(for FSSM-I)} \\ \\
	  M_S \lesssim  10^{8}  \text{GeV}  \\  \text{(for FSSM-II)}
         \end{cases} $
 & $\mu_\text{pole} \subset [1110,1140] $ GeV   &     \\[1em]
 \hline
\end{tabular}
\end{footnotesize}
\caption{ \footnotesize Approximate constraints on the SUSY scale and on pole masses for the Dark matter candidates. We impose a splitting between fake Higgsinos bigger than $300$ keV to avoid direct detection through inelastic scattering, we require a gluino life-time smaller than $100$ s to avoid hampering BBN and finally constrain the relic density (calculated at tree-level in \micrOMEGAs) to be $ \Omega h^2 \subset [0.1158 ,0.1218 ]$. When considering constraints on $M_S$, gaugino masses were taken in the multi-TeV range. }
\label{table:constraints}
\end{table}
The constraints on F-higgsinos dark matter (scenario $\mathbf{\HDM}$ ) are in particular quite stringent. Indeed, since their couplings to the Higgs and (F)-gauginos are suppressed, the neutral higgsinos have a very small splitting. They are therefore a perfect example of inelastic dark matter and direct detection experiments constrain their mass splitting $\delta$ to be bigger than roughly $300 $ keV~\cite{nagata_higgsino_2015,aprile_implications_2011,akerib_first_2014}. Estimating the splitting by 
\begin{align}
\label{eq:approxsplit}
 \delta & ~\simeq~  \begin{cases}
200 \ \text{keV} \cdot \scr{O}(1) \cdot \displaystyle \left( \frac{  400 \text{ TeV} }{ \MS}\right)^2   \left( \frac{  m_{fg} }{ 4 \text{ TeV}}\right) \quad \textrm{ for the FSSM-I}  \\
200 \ \text{keV} \cdot \scr{O}(1) \cdot \displaystyle \left( \frac{  10^7 \text{ GeV} }{ \MS}\right) \left(\frac{\mu }{ 1 \text{ TeV}}\right) \left( \frac{  4 \text{ TeV} }{ m_{fg}}\right) \quad \textrm{ for the FSSM-II}  \ ,
\end{cases} 
\end{align}
we recover the bounds from Table~\ref{table:constraints}.

\section{Conclusion}

Two topics have been discussed here.

Clearly, if future LHC experiments find light new fermions that can be identified with signals of a Split-SUSY scenario, an experimental investigation of their precise quantum numbers and couplings will need to be determined. FSSM give a framework and further motivation for conducting such investigation.

The other topic discussed deals with the propagation of a spin-3/2 in a non-Lorentz invariant background (the fundamentals laws remaining Lorentz invariant). The Lagrangian was motivated by the study of the super-Higgs mechanism in a fluid, but can be taken and studied per se. It exhibits a feature common with spin-3/2 propagating in a curved space: the longitudinal modes ($\pm 1/2$ helicities) propagate at a slower speed, hence the ``slow gravitino'' name.

\section*{Acknowledgments}

\noindent 
We would like to thank M.~D.~Goodsell, Y.~Oz, G. Policastro and P.~Slavich for 
%stimulating discussions that has lead to 
collaborations on the different subjects presented here.
This work is supported by the ERC advanced grant ERC Higgs@LHC, the ANR contract HIGGSAUTOMATOR and the Institut Lagrange de Paris.

\end{document}